\documentclass[prl,twocolumn,superscriptaddress,amsmath,amssymb,showpacs]{revtex4}
\pdfoutput=1

\usepackage{float,graphicx}
\newif\ifFIGS
\FIGStrue

\newcommand{\dof}{d.o.f.}
\newcommand{\oam}{OAM}
\newcommand\oaml{l}
\newcommand\oamr{r}
\newcommand\rsp{RSP}
\newcommand\resp{RESP}
\newcommand\bsa{BSA}

\newcommand\pbs{PBS}

\def\ketbra#1{\vert#1\rangle\langle#1\vert}
\def\openone{\leavevmode\hbox{\small1\kern-3.3pt\normalsize1}}

\def\ket#1{\vert#1\rangle}
\newcommand{\ktbr}[2]{|#1\rangle_{#2}\langle#1|}

\begin{document}

\title{Remote Preparation of Single-Photon ``Hybrid'' Entangled and\\ Vector-Polarization States} 
\author{Julio T. Barreiro} 
\affiliation{Department of Physics, University of
  Illinois at Urbana-Champaign, Urbana, Illinois 61801-3080, USA}
\affiliation{Institut f\"ur Experimentalphysik, Universit\"at
Innsbruck, Technikerstrasse 25, A-6020 Innsbruck, Austria}
\author{Tzu-Chieh Wei}
\affiliation{Department of Physics, University of
  Illinois at Urbana-Champaign, Urbana, Illinois 61801-3080, USA}
\affiliation{Department of Physics and Astronomy, University
  of British Columbia, Vancouver, British Columbia, Canada}
\author{Paul G. Kwiat} 
\affiliation{Department of Physics, University of
  Illinois at Urbana-Champaign, Urbana, Illinois 61801-3080, USA}

\begin{abstract}
Quantum teleportation faces increasingly demanding requirements for
transmitting large or even entangled systems.  However, knowledge of the state
to be transmitted eases its reconstruction, resulting in a protocol known as
remote state preparation.  A number of experimental demonstrations to date have
been restricted to single-qubit systems.  We report the remote preparation of
two-qubit ``hybrid'' entangled states, including a family of
vector-polarization beams.  Our single-photon states are encoded in the photon
spin and orbital angular momentum.  We reconstruct the states by spin-orbit
state tomography and transverse polarization tomography.  The high fidelities
achieved for the vector-polarization states opens the door to optimal coupling
of downconverted photons to other physical systems, such as an atom, as
required for scalable quantum networks, or plasmons in photonic nanostructures.
\end{abstract}


\maketitle 

Quantum communication involves the transfer of quantum information, either
directly by sending the quantum states or indirectly by using quantum and
classical resources.  The cost of indirect transfer involves a trade-off
between ebits, units of entanglement, and cbits, units of classical
communication, which depends on what is known about the state to the sender and
receiver.  For example, in teleportation~\cite{bennett-prl-70-1895}, an
\textit{unknown} quantum state is sent via a quantum channel, consuming 1 ebit,
and 2 bits of classical communication.  In contrast, if the state is
\textit{known} to the sender, the required resources can be reduced to 1 ebit
and 1 cbit, in a protocol named remote state preparation
(\rsp)~\cite{lo-pra-62-012313,pati-pra-63-014302,bennett-prl-87-077902,babichev-prl-92-047903}.
This variant of teleportation has received much attention lately given its
reduced resource requirements and known optimal schemes and
trade-offs~\cite{berry-prl-90-057901,abeyesinghe-pra-68-062319,killoran-pra-81-012334}.

In \rsp{}, Alice wishes to prepare a state in Bob's laboratory by relying on
the correlations of shared entangled states and classical communication.
\rsp{} not only requires fewer resources than teleportation, but also escapes
the need for Bell-state analysis (\bsa{}), impossible with linear optics, but
enabled by
hyperentanglement~\cite{kwiat-pra-58-R2623,schuck-prl-96-190501,barbieri-pra-75-042317,barreiro-nphys-4-282}.
\rsp{} has been realized for arbitrary one-qubit states with varying
efficiencies.  The first demonstration was 50\%
efficient~\cite{peters-prl-94-150502}, at the price of 1 ebit and 1 cbit, due
to the impossibility of a universal NOT operation on arbitrary qubit states.
Subsequent demonstrations achieved 100\% efficiency in principle, but required
as many resources as teleportation (1 ebit, 2
cbits)~\cite{rosenfeld-prl-98-050504,liu-pra-76-022308}; however, \bsa{} was
not necessary.

In this Letter, we show that by working in a larger Hilbert space, \rsp{} can
be extended to remotely prepare multiqubit states, including entangled
states~\cite{liu-pla-316-159}.  We implement this protocol with a cost of 2
ebits, and 2 cbits of forward classical communication (4 cbits for sending
completely arbitrary pure states, as shown below).  We then extend it to
prepare mixed states and a four-parameter family of pure states.  The latter
includes a remarkable family of states with nonuniform transverse
polarization, so called vector-polarization states~\cite{zhan-aop-1-1}.  These
states are important for their applications in improved
metrology~\cite{dorn-prl-91-233901}, ideal production of
plasmons~\cite{kano-josab-15-1381}, and in principle 100\%-efficient coupling
to an atom~\cite{sondermann-apb-89-489}.

Specifically, by using hyperentanglement, i.e., systems simultaneously
entangled in multiple degrees of
freedom~\cite{kwiat-jmo-44-2173,barreiro-prl-95-260501}, we remotely prepare
single-photon states entangled in their spin and orbital angular momentum
(\oam{})~\cite{barreiro-nphys-4-282}.  Such ``hybrid''
entanglement~\cite{boschi-prl-80-1121,zhao-prl-95-030502,ma-pra-79-042101,nagali-prl-103-013601,neves-pra-80-042322}
between the polarization and the spatial mode of a single photon can be easily
converted into a spatially separated single-particle state established to be
entangled~\cite{enk-pra-72-064306}.  In our remote entangled-state preparation
(\resp{}) protocol, Alice and Bob share a hyperentangled pair, e.g., a product
of Bell states of polarization and spatial modes
$\Phi^+_\text{spin}\otimes\Psi^+_\text{orbit}\equiv
\left(|HH\rangle+|VV\rangle\right)/\sqrt{2}\otimes
\left(|\oaml\oamr\rangle+|\oamr\oaml\rangle\right)/\sqrt{2}$, where $H$ ($V$)
represents the horizontal (vertical) photon polarization and $l$ ($r$)
represents the paraxial spatial mode (Laguerre-Gauss) carrying $+\hbar$
($-\hbar$) units of \oam~\cite{allen-2003-oam}.  For operations on individual
photons, we rewrite the shared state in the single-photon basis
as~\cite{barreiro-nphys-4-282}
\begin{equation}
\Phi^+_\text{spin}\otimes\Psi^+_\text{orbit} =%
\frac{1}{2}\big(\phi^+_A\otimes \psi^+_B + \phi^-_A\otimes\psi^-_B + %
\psi^+_A\otimes \phi^+_B + \psi^-_A\otimes \phi^-_B
\big),
\label{eq:respepr}
\end{equation}
where the single-photon ``spin-orbit'' states have the Bell-state form:
$\phi^\pm \equiv \frac{1}{\sqrt{2}}(|H\oaml\rangle\pm|V\oamr\rangle),\quad
\psi^\pm \equiv \frac{1}{\sqrt{2}}(|H\oamr\rangle\pm|V\oaml\rangle)$.  Thus,
when Alice measures her photon (A) with a spin-orbit \bsa{}, the state of Bob's
photon (B) is projected into one of the four spin-orbit entangled states
$\phi_B^\pm$, $\psi_B^\pm$, according to Alice's outcome $\psi_A^\pm$,
$\phi_A^\pm$.  Alice can remotely prepare Bob's single photon into a desired
spin-orbit entangled state by letting him know the correcting unitary
transformation in 2 cbits; e.g., to prepare $\psi_B^+$, Alice tells Bob to do
nothing if her outcome is $\phi_A^+$, transform $V\rightarrow-V$ for
$\phi_A^-$, $H\leftrightarrow V$ for $\psi_A^+$, or $V\rightarrow-V$ and
$H\leftrightarrow V$ for $\psi_A^-$.

\begin{figure}
\ifFIGS\centerline{\includegraphics{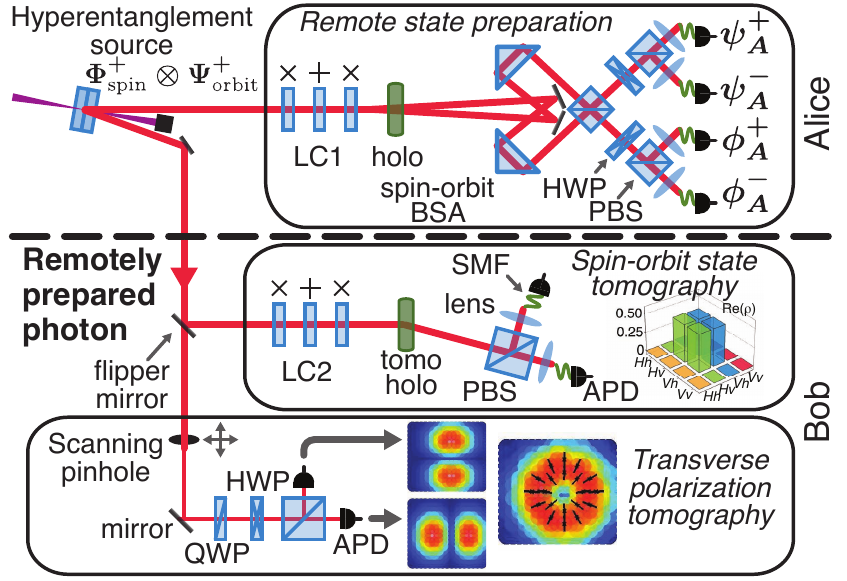}}\fi
\caption{Experimental setup for the remote preparation of single-photon
  entangled and vector-polarization states.  LC, liquid crystals (LC1 for state
  preparation, LC2 for polarization tomography) with optic axis perpendicular
  to the beam and oriented as indicated ($+,\times$); BSA, Bell-state analyzer;
  QWP, quarter-wave plate; HWP, Half-wave plate; SMF, single-mode fiber; APD,
  avalanche photodiode; PBS, polarizing beam splitter; holo, forked
  binary-grating hologram; tomo holo, holograms for spatial-mode
  tomography~\cite{barreiro-prl-95-260501}.}
\label{fig:setup}
\end{figure}

\begin{figure}
\ifFIGS\centerline{\includegraphics{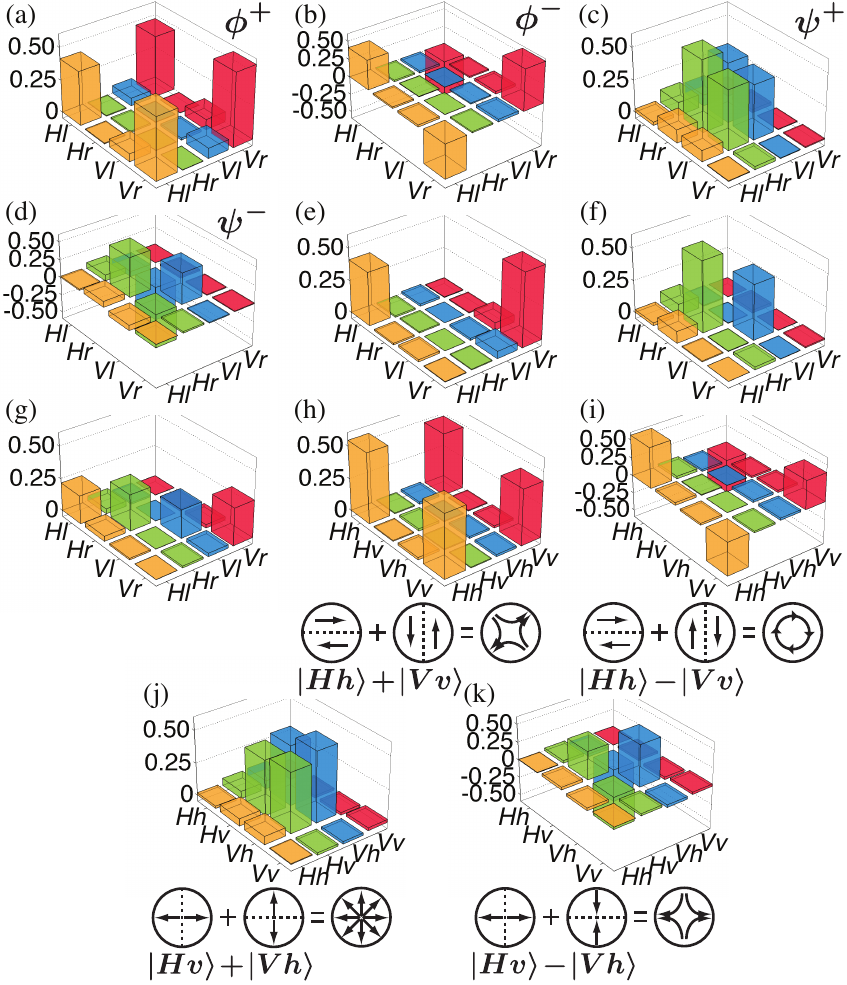}}\fi
\caption{Experimental density matrices (real parts) of remotely prepared
  single-photon two-qubit states.  (a)-(d) Maximally entangled spin-orbit
  states, (e),(f) partially mixed states, and (g) completely mixed state.
  Family of vector polarization states, (h)-(k), with their ideal polarization
  profiles shown underneath [spatial-mode component of the density matrix shown
  in the basis $|h/v\rangle=(|l\rangle\pm|r\rangle)/\sqrt{2}$].  The average
  magnitude of all imaginary elements, not shown, is 0.02.}
\label{fig:resp}
\end{figure}

We implemented this \resp{} protocol using our spin-orbit \bsa{} and
tomographically reconstructed the remotely prepared states
(Fig.~\ref{fig:setup}; for details on our source and spin-orbit \bsa{},
see~\cite{barreiro-nphys-4-282,epaps}).  The tomographic measurements are
similar to those used for hyperentangled states~\cite{barreiro-prl-95-260501},
but in this case applied only to Bob's photon.  The spin-orbit state tomography
consists of polarization tomography, performed by liquid crystals and a
polarizing beam splitter (\pbs{}), and spatial-mode tomography, realized by
mode-transforming holograms and single-mode fibers.  Figures~\ref{fig:resp}a-d
show the reconstructed density matrices of the four canonical, remotely
prepared, spin-orbit Bell states.  The high quality of the prepared states is
captured in Table~\ref{tab:quality}, where we quote their fidelity with the
target state, degree of entanglement (tangle), and mixture (linear
entropy)~\cite{white-josab-24-172}.

\begin{table}
  \caption{Quality parameters of remotely prepared two-qubit states.  The first
    column identifies the states in Fig.~\ref{fig:resp}.  Other columns show
    each prepared-state fidelity ($F$) with the target state, tangle ($T$), and
    linear entropy ($S_L$).  For mixed states we used longer acquisition times,
    leading to smaller uncertainties.  The partially mixed states [2(e) and
      2(f)] have an ideal $S_L=2/3$ and $T=0$; all others have $S_L=0$ and
    $T=1$, except the completely mixed state 2(g), for which $S_L=1$ and $T=0$.
    All errors are calculated from Monte Carlo simulations of Poissonian
    counting statistics.}
\begin{ruledtabular}
\begin{tabular}{ccccc}
Figure&Target state & $F$ & $T$ & $S_L$ \\\hline
2(a)&$\phi^+$&$0.955(2)$&$0.86(1)$&$0.06(1)$\\
2(b)&$\phi^-$&$0.968(2)$&$0.90(1)$&$0.06(1)$\\
2(c)&$\psi^+$&$0.938(3)$&$0.85(2)$&$0.07(1)$\\
2(d)&$\psi^-$&$0.949(3)$&$0.85(1)$&$0.08(1)$\\
2(e)&$\left(\ketbra{Hl}+\ketbra{Vr}\right)/2$&$0.967(2)$&$0.005(2)$&$0.666(3)$\\
2(f)&$\left(\ketbra{Hr}+\ketbra{Vl}\right)/2$&$0.961(3)$&$0.015(3)$&$0.658(3)$\\
2(g)&$\frac{1}{4}\openone$&$0.986(1)$&$0.000(1)$&$0.982(1)$\\
2(h)&$\left(|Hh\rangle+|Vv\rangle\right)/\sqrt{2}$&$0.964(3)$&$0.88(1)$&$0.07(1)$\\
2(i)&$\left(|Hh\rangle-|Vv\rangle\right)/\sqrt{2}$&$0.940(5)$&$0.82(2)$&$0.06(1)$\\
2(j)&$\left(|Hv\rangle+|Vh\rangle\right)/\sqrt{2}$&$0.938(3)$&$0.82(1)$&$0.10(1)$\\
2(k)&$\left(|Hv\rangle-|Vh\rangle\right)/\sqrt{2}$&$0.928(3)$&$0.81(1)$&$0.12(1)$
\end{tabular}
\end{ruledtabular}
\label{tab:quality}
\end{table}

Note that the spin and orbit \dof{} of Bob's particle become entangled without
local interaction; the above protocol thus realizes entanglement swapping.
However, instead of swapping the entanglement between two pairs of particles,
here the swapping occurs between pairs of degrees of freedom.  The high quality
of our scheme (measured fidelities $\approx 95\%$), compares favorably to the
best reported for entanglement swapping~\cite{riebe-nphys-4-839}.

In order to use the protocol to remotely prepare \emph{mixed} states, Alice
needs to induce decoherence on Bob's photon.  One way to achieve this is by
entangling the quantum system to yet another of its \dof{}, which is then
traced over.  This technique has previously enabled the precise remote
preparation of single-qubit mixed states~\cite{peters-prl-94-150502}.  There,
Alice coupled her photon's polarization and frequency \dof{} followed by a
frequency-insensitive measurement, thus preparing Bob's photon in a mixed
state.  Similarly, for \resp{}, Alice can couple the \oam{} and frequency
\dof{} by detuning the spin-orbit \bsa{} interferometer.  After a
frequency-insensitive measurement, Alice measures spin-orbit mixed states,
preparing Bob's photon in a controllable spin-orbit mixed state.

Alternatively, we can trace over Alice's photon spin and/or orbital \dof{} or a
subspace of them.  For example, consider the spin-orbit \bsa{} without either
half-wave plates (HWPs) or PBSs (see Fig.~\ref{fig:setup}).  In this case, instead of the
outcomes $\phi^+_A$ and $\phi^-_A$ ($\psi^+_A$ and $\psi^-_A$) we only have the
outcome $\phi_A$ ($\psi_A$).  Consequently, when a pair in the state of
Eq.~\ref{eq:respepr} is shared and Alice detects a photon in $\phi_A$
($\psi_A$), Bob's photon is prepared in the classically correlated state
$\sim\ktbr{\psi^+}{B}+\ktbr{\psi^-}{B}=\ketbra{Hr}+\ketbra{Vl}$ or
$\sim\ktbr{\phi^+}{B}+\ktbr{\phi^-}{B}=\ketbra{Hl}+\ketbra{Vr}$, at the cost of
1 cbit.  In addition, by acting on Alice's polarization (as discussed below),
we can also prepare classically correlated states in other bases.  Furthermore,
if both spin and \oam{} of Alice's photon are ignored, Bob's photon is left in
a two-qubit completely mixed state (without classical communication).  We
efficiently prepared such classically correlated and completely mixed states,
whose reconstructed density matrices are shown in Figs.~\ref{fig:resp}e-g.

We can readily remotely prepare a particularly interesting four-parameter
family of states by acting on Alice's photon polarization and implementing a
modified spin-orbit \bsa{}.  Alice and Bob initially share the hyperentangled
state $\Phi^+_\text{pol}\otimes\Psi^+_\text{spa}$.  Alice then applies to her
photon the polarization unitary operation: $\ket{H}\longrightarrow \cos \theta
\ket{H} + e^{i\eta}\sin \theta \ket{V}$, $\ket{V}\longrightarrow e^{i\phi}(\sin
\theta \ket{H} - e^{i\eta}\cos \theta \ket{V})$; or
$e^{i(\eta+\phi)/2}R_z(\eta) R_y(2\theta) R_z(\phi)$, in terms of the Bloch
rotation operators, implemented using the LC1 liquid crystals (see
Fig.~\ref{fig:setup}).  Next, Alice measures her photon with a ``rotated''
spin-orbit \bsa{}:
\begin{eqnarray*}
\phi^+_A(\alpha) \equiv \cos\alpha\ket{Hr}_A +
\sin\alpha\ket{Vl}_A\\ \phi^-_A(\alpha) \equiv \sin\alpha\ket{Hr}_A -
\cos\alpha\ket{Vl}_A\\ \psi^+_A(\beta) \equiv \cos\beta\ket{Hl}_A +
\sin\beta\ket{Vr}_A\\ \psi^-_A(\beta) \equiv \sin\beta\ket{Hl}_A -
\cos\beta\ket{Vr}_A.
\end{eqnarray*}
Such a measurement consists of a spin-orbit \bsa{} in which the last
polarization measurement is made at the angle $\alpha$ ($\beta$) instead of
$45^\circ$ for the $\phi^\pm$ ($\psi^\pm$) output; experimentally, $\alpha$ and
$\beta$ are set with HWPs before the final \pbs{} in Alice's setup, as shown in
Fig.~\ref{fig:setup}.

Taking into account the unitary polarization operation and the ``rotated''
spin-orbit \bsa{}, we can rewrite the shared state as
\begin{eqnarray*}
\Phi^+_\text{spin}\otimes\Psi^+_\text{oam}\longrightarrow
\phi^+_A(\alpha)\otimes\phi^+_B(\alpha,\eta,\theta,\phi)+\\
\phi^-_A(\alpha)\otimes\phi^-_B(\alpha,\eta,\theta,\phi)+\\
\psi^+_A(\beta)\otimes\psi^+_B(\beta,\eta,\theta,\phi)+\\
\psi^-_A(\beta)\otimes\psi^-_B(\beta,\eta,\theta,\phi),\phantom{+}
\end{eqnarray*}
where Alice's possible outcomes $\phi^\pm_A(\alpha)$ or $\psi^\pm_A(\beta)$
determine Bob's state:
\begin{align*}
\phi^+_B(\alpha,\eta,\theta,\phi) &\equiv \cos\alpha\ket{\Xi,r} + e^{i\eta}\sin\alpha\ket{\Xi^\perp,l}\\ 
\phi^-_B(\alpha,\eta,\theta,\phi) &\equiv \sin\alpha\ket{\Xi,r} - e^{i\eta}\cos\alpha\ket{\Xi^\perp,l}\\
\psi^+_B(\beta,\eta,\theta,\phi) &\equiv \cos\beta\ket{\Xi,l} + e^{i\eta}\sin\beta\ket{\Xi^\perp,r}\\
\psi^-_B(\beta,\eta,\theta,\phi) &\equiv \sin\beta\ket{\Xi,l} -
e^{i\eta}\cos\beta\ket{\Xi^\perp,r},
\end{align*}
with $\ket{\Xi} \equiv \cos\theta\ket{H} - e^{i\phi}\sin\theta\ket{V}$ and
$\ket{\Xi^\perp} \equiv \sin\theta\ket{H} + e^{i\phi}\cos\theta\ket{V}.$

The states in this four-parameter family have the remarkable property that their
transverse polarization profiles are not in general uniform~\cite{maurer-njp-9-78}.
Of outstanding interest is the family of states
$\frac{1}{\sqrt{2}}\left(\ket{Rr}\pm\ket{Ll}\right)$ and
$\frac{1}{\sqrt{2}}\left(\ket{Rl}\pm\ket{Lr}\right)$, because of their
potentially useful polarization profiles [see Figs.~\ref{fig:resp}(h)-(k)].  In
particular, states with radial polarization
profiles~\cite{zhan-aop-1-1}, such as
$\ket{Rr}-\ket{Ll}=\ket{Hv}+\ket{Vh}$, have enabled a focused spot size
significantly smaller than possible with linear
polarization~\cite{dorn-prl-91-233901}.  Theoretically, such states have also
been shown to enable the largest possible longitudinal electric field component
in the focal point of a lens~\cite{urbach-prl-100-123904}, leading to optimal
coupling to plasmons in subwavelength-diameter holes~\cite{kano-josab-15-1381}.
Additionally, the radial polarization state is predicted to enable 100\%
light-atom coupling in free space~\cite{sondermann-apb-89-489}, albeit with a
somewhat different radial amplitude distribution.

We remotely prepared a variety of examples from this family of states.  Alice's
liquid crystals LC1 implement the required polarization unitary,
$\ket{H_A}\rightarrow(\ket{H_A}+\ket{V_A})/\sqrt{2},\;\ket{V_A}\rightarrow
i(\ket{H_A}-\ket{V_A}) /\sqrt{2}$.  After measuring Alice's photon with a
spin-orbit \bsa{}, we tomographically reconstructed the states of Bob's photon,
resulting in the density matrices shown in Fig.~\ref{fig:resp}h-k.

\begin{figure}
\ifFIGS\centerline{\includegraphics{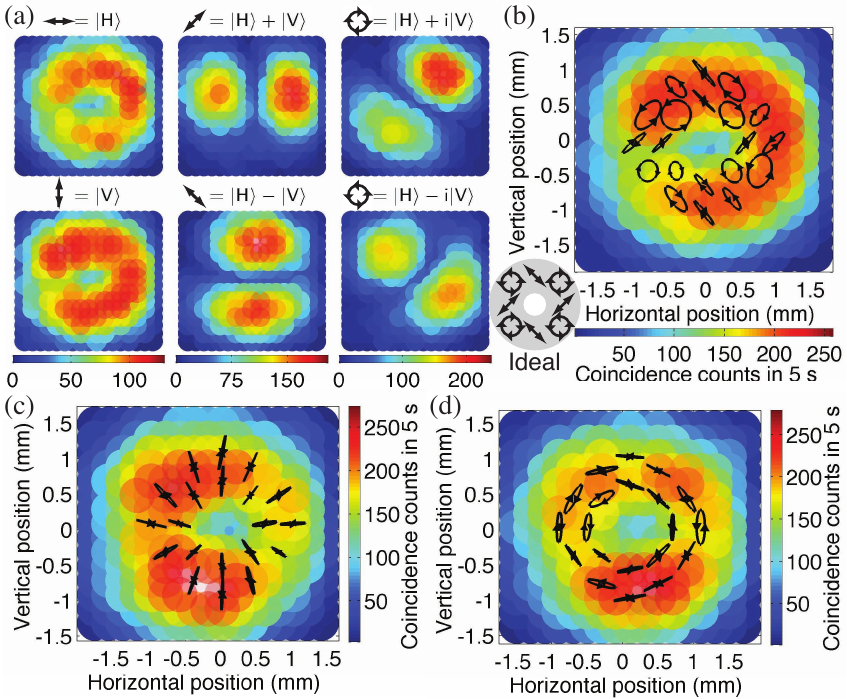}}\fi
\caption{Transverse polarization and intensity profiles of remotely prepared
  vector beams.  For a canonical maximally entangled spin-orbit state $\phi^-
  \equiv \frac{1}{\sqrt{2}}(|H\oaml\rangle-|V\oamr\rangle)$, we show (a), the
  intensity profiles for each polarization projection, and (b), the
  polarization profiles from state tomography at each point; the average
  fidelity with the target states over all sampled points is
  $F_\text{av}=95(4)\%$.  (c), Radial polarization, $F_\text{av}=94(4)\%$,
  and (d), azimuthal polarization, $F_\text{av}=95(2)\%$.  The polarization
  ellipses are shown for a subset of all samples and their size suggests the
  magnitude of state purity, where pure states have the size of the colored
  circles (equivalent to the size of the collection pinhole).}
\label{fig:profiles}
\end{figure}

In order to directly verify the vector profile of the remotely prepared modes,
we also implemented a direct transverse polarization tomography.  On Bob's
beam, at a location with a beam waist size of 1.15(5)mm (measured with the
Gaussian component of the down-converted beam), we transversely scanned a
500-$\mu$m diameter pinhole in 200-$\mu$m steps.  At each point in a $16\times
16$ grid, we performed a polarization tomography (6 measurements, 5-sec
acquisition time for each).  The results of the reconstruction are shown in
Fig.~\ref{fig:profiles}.  We found the expected position of the center of the
beam by maximizing the overlap between our measurements and those expected for
an ideal beam with our measured waist.  As shown in Fig.~\ref{fig:profiles}, we
achieve a high average fidelity of $\approx95\%$, over all pinhole positions,
between the measured and ideal polarizations.

Having discussed the remote preparation of a specific four-parameter family of
single-photon states, we can now ask: Can Alice remotely prepare
\emph{arbitrary} pure states $\Psi_B(a,b,c,d)=a\,\phi^+_B + b\,\phi^-_B +
c\,\psi^+_B + d\,\psi^-_B$ ($a$-$d$ complex)?  The answer is indeed ``yes'',
via the use of positive operator value measure, as described
online~\cite{epaps}.  Furthermore, to remotely prepare arbitrary mixed states,
Alice can prepare the ensemble of pure states $\{p_i,\psi_i\}$ so that Bob
receives $\rho_B=\sum_i p_i |\psi_{B,i}\rangle\langle \psi_i|$.

In summary, we presented the first demonstration of a powerful new technique to
remotely prepare a wide variety of single-photon entangled states, using the
resource of hyperentangled photon pairs.  Some of the states prepared are
already known to be optimal for several applications~\cite{zhan-aop-1-1}, and
we anticipate other uses to be revealed by further investigation.  A radial
polarization state, for example, could enable an optimal plasmon-assisted
transmission of entangled photons, remarkably improving earlier demonstrations
using entanglement in a single degree of freedom
(polarization~\cite{altewischer-nature-418-304},
energy-time~\cite{fasel-prl-94-110501}, and \oam{}~\cite{ren-el-76-753}).
Furthermore, we have shown how the protocol may be modified to allow the remote
preparation of arbitrary two-qubit states.  It will be interesting to consider
the generalization to include other degrees of freedom as well, to explore an
even richer space of states.

We acknowledge helpful discussions with Nicholas Peters, and funding support
from the ADNA/S\&T-IARPA project Hyperentanglement-Enhanced Advanced Quantum
Communication (NBCHC070006) and NSF Grant No. PHY-0903865.

\end{document}


\title{Remote preparation of single-photon ``hybrid'' entangled and\\
  vector-polarization states\\ SUPPLEMENTAL MATERIAL}
 
\author{Julio T. Barreiro} 
\affiliation{Department of Physics, University of
  Illinois at Urbana-Champaign, Urbana, Illinois 61801-3080, USA}
\affiliation{Institut f\"ur Experimentalphysik, Universit\"at
Innsbruck, Technikerstrasse 25, A-6020 Innsbruck, Austria}
\author{Tzu-Chieh Wei}
\affiliation{Department of Physics, University of
  Illinois at Urbana-Champaign, Urbana, Illinois 61801-3080, USA}
\affiliation{Department of Physics and Astronomy, University
  of British Columbia, Vancouver, British Columbia, Canada}
\author{Paul G. Kwiat} 
\affiliation{Department of Physics, University of
  Illinois at Urbana-Champaign, Urbana, Illinois 61801-3080, USA}


\maketitle 

\section{Source}  

The hyperentanglement source is realized by directing 120 mW of 351-nm light
from a continuous-wave (CW) Ar$^+$ laser into two contiguous $\beta$-barium
borate (BBO) nonlinear crystals with optic axes aligned in perpendicular
planes~[J.~T.~Barreiro, T.-C.~Wei, and P.~G.~Kwiat, Nature Physics
  \textbf{4}, 282 (2008)].  Type-I degenerate 702-nm photons in a 3.0$^\circ$
    half-opening angle cone are produced by phase-matching each 0.6-mm thick
    crystal.  In the spin and $\pm1$ \oam{} subspace, a two-fold coincidence
    rate of 5 detected pairs/s is determined by a 10 ns coincidence window and
    interference filters with $\Delta\lambda_\text{FWHM}=5\;$nm.

\section{Arbitrary two-qubit RESP}

Alice can remotely prepare \emph{arbitrary} pure states
$\Psi_B(a,b,c,d)=a\,\phi^+_B + b\,\phi^-_B + c\,\psi^+_B + d\,\psi^-_B$
($a$-$d$ complex) as follows.  Alice determines the positive operator-valued
measure (\povm{}) constructed by the following Kraus operators,
\begin{eqnarray}
F_1&=&a|\phi^+\rangle_A\langle|+b|\phi^-\rangle_A\langle|+c|\psi^+\rangle_A\langle|+d|\psi^-\rangle_A\langle|\nonumber\\
F_2&=&b|\phi^+\rangle_A\langle|+c|\phi^-\rangle_A\langle|+d|\psi^+\rangle_A\langle|+a|\psi^-\rangle_A\langle|\nonumber\\
F_3&=&c|\phi^+\rangle_A\langle|+d|\phi^-\rangle_A\langle|+a|\psi^+\rangle_A\langle|+b|\psi^-\rangle_A\langle|\nonumber\\
F_4&=&d|\phi^+\rangle_A\langle|+a|\phi^-\rangle_A\langle|+b|\psi^+\rangle_A\langle|+c|\psi^-\rangle_A\langle|,\nonumber
\end{eqnarray}
where $|\chi\rangle\langle|\equiv|\chi\rangle\langle\chi|$.  By normalization
of the state, $|a|^2+|b|^2+|c|^2+|d|^2=1$, they satisfy the completeness relation
$\sum_i F^\dagger_i F_i=\openone$.

Now suppose, e.g., that Alice wants to prepare $\Psi_B(a,b,c,d)$ and her
\povm{} measurement gives the outcome $F_3$.  The joint state in
Eq.~(1) in the Letter then becomes
\begin{equation*}
a\,\psi^+_A\otimes\phi^+_B + b\,\psi^-_A\otimes\phi^-_B + 
c\,\phi^+_A\otimes\psi^+_B + d\,\phi^-_A\otimes\psi^-_B.
\end{equation*}
Next, Alice performs a von Neumann measurement on the filtered state in the
basis of $(i,ii) {\sim\left(|H\rangle\pm |V\rangle\right)|\oaml\rangle_A}$,
$(iii,iv) {\sim\left(|H\rangle\pm |V\rangle\right)|\oamr\rangle_A}$ (\eg, with
an \oam{} splitter, a HWP@45$^\circ$ and a \pbs{}).  If the outcome is $(i)$,
Bob's photon will be projected into the state $\Psi_B(a,b,c,d)=a \phi^+_B + b
\phi^-_B+ c \psi^+_B + d\psi^-_B$. Similarly, the other outcomes $(ii)-(iv)$
give $\Psi_B(-a,b,c,d)$, $\Psi_B(a,b,c,-d)$, and $\Psi_B(a,b,-c,d)$,
respectively, i.e. $\Psi_B(a,b,cd)$ with a phase error; Bob can correct each of
these with 2 bits of information from Alice.  Furthermore, for the other
\povm{} outcomes $F_{i\neq3}$, the state is a permutation of the spin-orbit
states, and again Alice can send 2 bits to Bob informing him how to correct the
state.

Thus, the protocol can remotely prepare any spin-orbit state $\Psi_B(a,b,c,d)$
with 4 cbits and 2 ebits.
Naturally, a single classical bit suffices if ones accepts a reduced
probability of success, i.e., Alice simply only sends a ``1'' (keep the state)
to Bob if she measures the desired state.